\documentclass[12pt]{article}
\usepackage{graphicx}

\begin{document}
\pagenumbering{arabic}
\begin{titlepage}
\title{Gravitational pressure on event horizons and thermodynamics
in the teleparallel framework}

\author{J. W. Maluf$\,^{(1,a)}$, S. C. Ulhoa$\,^{(2,b)}$ 
and J. F. da Rocha-Neto$\,^{(1,c)}$}

\date{}
\maketitle

{\footnotesize
\noindent{  (1) Instituto de F\'{\i}sica, Universidade de 
Bras\'{\i}lia, C.P. 04385, 70.919-970 Bras\'{\i}lia DF, Brazil}\par
\bigskip
\noindent{ (2) Faculdade UnB Gama, Universidade de Bras\'ilia,
72.405-610, Gama DF, Brazil}}
\bigskip

\begin{abstract}
The concept of gravitational pressure is naturally defined in the context of
the teleparallel equivalent of general relativity. Together with the definition
of gravitational energy, we investigate the thermodynamics of rotating black
holes in the teleparallel framework.
We obtain the value of the gravitational pressure over the external event
horizon of the Kerr black hole, and write an expression for the 
thermodynamic relation $TdS =dE + pdV$, where the variations refer to the 
Penrose process for the Kerr black hole. We employ only the notions of 
gravitational energy and pressure that arise in teleparallel gravity, and  
do not make any consideration of the area or the variation of the area of 
the event horizon. However, our results are qualitatively similar to the 
standard expression of the literature.
\end{abstract}
\vfill
\par

\bigskip
{\footnotesize
\noindent (a) wadih@unb.br, jwmaluf@gmail.com\par
\noindent (b) sc.ulhoa@gmail.com\par
\noindent (c) rocha@fis.unb.br\par}

\end{titlepage}
\bigskip
\section{Introduction}
The discovery by Hawking \cite{Hawking} of the result that the area $A$ of an
arbitrary black hole never decreases, together with the idea of the Penrose
process \cite{Penrose} of extraction of energy of rotating black holes, led to
the establishment of the thermodynamics of the gravitational field. The 
analysis in the literature is normally restricted to the geometrical and 
dynamical behaviour of 
the area of the external event horizon of rotating black holes. By means of
the Penrose process the initial mass $m$ and angular momentum $J$ of 
the Kerr black hole vary by $dm$ and $dJ$, respectively, such 
that $dm -\Omega_H dJ >0$, where $\Omega_H$ is 
the angular velocity of the external event horizon of the black hole. In 
this process, the variation of the area $A$ of the black hole satisfies 
$dA > 0$. In the final stage of an idealized Penrose process, the mass of the 
black hole becomes the irreducible mass $m_{irr}$ \cite{CR} defined by the 
relation $m^2=m_{irr}^2+J^2/(4m_{irr}^2)$.

Let us denote, as usual, the parameter $a$ as the angular momentum per unit 
mass of the Kerr black hole, $a=J/m$, and $S$ and $T$ as the gravitational 
entropy and temperature, respectively. The ordinary identification of the
gravitational thermodynamic quantities with  the parameters $(m,a)$ of the
Kerr space-time is the following: $S \rightarrow A/4$ and 
$T\rightarrow k_s/(2\pi)$; $k_s$ is the
surface gravity defined by $k_s=\sqrt{m^2-a^2} /(2mr_+)$, and 
$r_+=m+\sqrt{m^2-a^2}$ is the radius of the external event horizon of the black
hole. The ordinary thermodynamic relation for the Kerr black hole reads

\begin{equation}
TdS= dm -\Omega_H dJ\,.
\label{1}
\end{equation}

In this paper, we will obtain an expression for the thermodynamic relation
$TdS =dE + pdV$ in the context of the teleparallel equivalent of general
relativity (TEGR). The expression for the gravitational energy $E$ is well
defined in the TEGR. We will show in the present analysis that the notion of
gravitational pressure $p$ over the external event horizon of the black hole is
also well defined, and naturally arises from the field equations and from the 
gravitational energy-momentum tensor defined in the realm of the TEGR. The 
spatial components of the energy-momentum (or stress-energy) tensor yield the
standard definition of the gravitational pressure, as we will explain.
The variation $dV$ is obtained by means of the variation of $r_+$, when the
parameters $m$ and $a$ of the black hole vary by the amounts $dm$ and $dJ$
($p$ turns out to be a density, and therefore $dV$ is actually given by
$dV=dr d\theta d\phi$).
We will show that our result for $TdS = dE + pdV$ is qualitatively and 
strikingly similar to eq. (1).  We will restrict our analysis to the 
external event horizon of the Kerr black hole (defined by $r=r_+$), but 
will not consider any property of the area $A$ (or $dA$) of the horizon.
The indications are that the efficiency  of the Penrose process in the present
context is
lower than in the ordinary thermodynamic formulation. We mention that the
use of the concept of pressure in the first law of black hole thermodynamics 
has been made in two recent investigations \cite{Dolan}, by considering that
the cosmological constant plays the role of pressure.

In section 2 we present the Lagrangian formulation of the TEGR and indicate 
how the definitions of the gravitational energy-momentum $P^a$, of the 
gravitational energy-momentum tensor $t^{\lambda \mu}$ and of the gravitational
pressure arise from the field equations of the theory. In section
3 we apply the definition of the gravitational pressure to the
external event horizon of the Kerr black hole. In section 4 we obtain the 
expression for $TdS$ that emerges in the present framework, and 
compare it with eq. (1). Finally, in section 6 we present our conclusions.

\bigskip
Notation: space-time indices $\mu, \nu, ...$ and SO(3,1) indices $a, b, ...$
run from 0 to 3. Time and space indices are indicated according to
$\mu=0,i,\;\;a=(0),(i)$. The tetrad field is denoted $e^a\,_\mu$, and the 
torsion tensor reads $T_{a\mu\nu}=\partial_\mu e_{a\nu}-\partial_\nu e_{a\mu}$.
The flat, Minkowski spacetime metric tensor raises and lowers tetrad indices 
and is fixed by $\eta_{ab}=e_{a\mu} e_{b\nu}g^{\mu\nu}= (-1,+1,+1,+1)$. The 
determinant of the tetrad field is represented by $e=\det(e^a\,_\mu)$.
\bigskip

\section{The gravitational energy-momentum and pressure in the TEGR}

In this section, we present a brief summary of the formulation of the 
teleparallel equivalent of general relativity. In the TEGR
the gravitational field 
is represented by the tetrad field $e^a\,_\mu$ only, and the Lagrangian density
is written in terms of the torsion tensor 
$T_{a\mu\nu}=\partial_\mu e_{a\nu}-\partial_\nu e_{a\mu}$. This tensor is 
related to the antisymmetric part of the Weitzenb\"ock connection 
$\Gamma^\lambda_{\mu\nu}=e^{a\lambda}\partial_\mu e_{a\nu}$. However, the
dynamics of the gravitational field in the TEGR is essentially the same as in
the usual metric formulation. The physics in both formulations is the same.

We first consider the torsion-free, Levi-Civita connection $^0\omega_{\mu ab}$,

\begin{eqnarray}
^0\omega_{\mu ab}&=&-{1\over 2}e^c\,_\mu(
\Omega_{abc}-\Omega_{bac}-\Omega_{cab})\,, \\ \nonumber
\Omega_{abc}&=&e_{a\nu}(e_b\,^\mu\partial_\mu
e_c\,^\nu-e_c\,^\mu\partial_\mu e_b\,^\nu)\,.
\label{2}
\end{eqnarray}
The Christoffel symbols ${}^0\Gamma^\lambda_{\mu\nu}$ and the
Levi-Civita connection are identically related by 

$$^0\Gamma^\lambda_{\mu\nu}=e^{a\lambda}\partial_\mu e_{a\nu}+
e^{a\lambda}\,(^0\omega_{\mu ab})e^b\,_\nu\,.$$ 
In view of this expression, an identity arises between the Levi-Civita 
connection and the contorsion tensor $K_{\mu ab}$, 

\begin{equation}
^0\omega_{\mu ab}=-K_{\mu ab}\,, 
\label{3}
\end{equation}
where $K_{\mu ab}=\frac{1}{2}e_{a}\,^{\lambda}e_{b}\,^{\nu}
(T_{\lambda\mu\nu}+T_{\nu\lambda\mu}+T_{\mu\lambda\nu})$, and 
$T_{\lambda\mu\nu}= e^a\,_\lambda T_{a\mu\nu}$.
The identity given by eq. (3) may be used to show that the scalar
curvature $R(e)$ may be identically written as

\begin{equation}
eR(^0\omega) =-e\left({1\over 4}T^{abc}T_{abc}+{1\over
2}T^{abc}T_{bac}-T^aT_a\right) +2\partial_\mu(eT^\mu)\,,
\label{4}
\end{equation}
where $e$ is the determinant of the tetrad field. Therefore in the framework
of the TEGR the Lagrangian density for the gravitational and matter fields 
reads 

\begin{eqnarray}
L&=& -k e({1\over 4}T^{abc}T_{abc}+{1\over 2}T^{abc}T_{bac}-
T^aT_a) -{1\over c}L_M \nonumber \\
&\equiv& -ke\Sigma^{abc}T_{abc} -{1\over c}L_M\,, 
\label{5}
\end{eqnarray}
where $k=c^3/16\pi G$, $T_a=T^b\,_{ba}$, 
$T_{abc}=e_b\,^\mu e_c\,^\nu T_{a\mu\nu}$ and

\begin{equation}
\Sigma^{abc}={1\over 4} (T^{abc}+T^{bac}-T^{cab})
+{1\over 2}( \eta^{ac}T^b-\eta^{ab}T^c)\;.
\label{6}
\end{equation}
$L_M$ stands for the Lagrangian density for the matter fields. 
The Lagrangian density $L$ is invariant under the global SO(3,1) group. The 
absence in the Lagrangian density of the divergent term on the right hand
side of eq. (4) prevents the invariance of (5) under arbitrary local SO(3,1)
transformations.

The field equations derived from (5) are equivalent to Einstein's equations. 
They are given by

\begin{equation}
e_{a\lambda}e_{b\mu}\partial_\nu (e\Sigma^{b\lambda \nu} )-
e (\Sigma^{b\nu}\,_aT_{b\nu\mu}-
{1\over 4}e_{a\mu}T_{bcd}\Sigma^{bcd} )={1\over {4kc}}eT_{a\mu}\,,
\label{7}
\end{equation}
where
$\delta L_M / \delta e^{a\mu}=eT_{a\mu}$. From now on we will make $c=1=G$.

The definition of the gravitational energy-momentum may be established in the
framework of the Lagrangian formulation defined by (5), according to the 
procedure of ref. \cite{Maluf1}. Equation (7) may be rewritten as 

\begin{equation}
\partial_\nu(e\Sigma^{a\lambda\nu})={1\over {4k}}
e\, e^a\,_\mu( t^{\lambda \mu} + T^{\lambda \mu})\;,
\label{8}
\end{equation}
where $T^{\lambda\mu}=e_a\,^{\lambda}T^{a\mu}$ and
$t^{\lambda\mu}$ is defined by

\begin{equation}
t^{\lambda \mu}=k(4\Sigma^{bc\lambda}T_{bc}\,^\mu-
g^{\lambda \mu}\Sigma^{bcd}T_{bcd})\,.
\label{9}
\end{equation}
In view of the antisymmetry property 
$\Sigma^{a\mu\nu}=-\Sigma^{a\nu\mu}$, it follows that

\begin{equation}
\partial_\lambda
\left[e\, e^a\,_\mu( t^{\lambda \mu} + T^{\lambda \mu})\right]=0\,.
\label{10}
\end{equation}
The equation above yields the continuity (or balance) equation,

\begin{equation}
{d\over {dt}} \int_V d^3x\,e\,e^a\,_\mu (t^{0\mu} +T^{0\mu})
=-\oint_S dS_j\,
\left[e\,e^a\,_\mu (t^{j\mu} +T^{j\mu})\right]\,.
\label{11}
\end{equation}
Therefore we identify
$t^{\lambda\mu}$ as the gravitational energy-momentum tensor \cite{Maluf1},

\begin{equation}
P^a=\int_V d^3x\,e\,e^a\,_\mu (t^{0\mu} 
+T^{0\mu})\,,
\label{12}
\end{equation}
as the total energy-momentum contained in a volume $V$ of the 
three-dimensional space,

\begin{equation}
\Phi^a_g=\oint_S dS_j\,
\, (e\,e^a\,_\mu t^{j\mu})\,,
\label{13}
\end{equation}
as the gravitational energy-momentum flux \cite{Maluf2}, and

\begin{equation}
\Phi^a_m=\oint_S dS_j\,
\,( e\,e^a\,_\mu T^{j\mu})\,,
\label{14}
\end{equation}
as the energy-momentum flux of matter. In view of (8) eq. (12) may be written
as 

\begin{equation}
P^a=-\int_V d^3x \partial_j \Pi^{aj}=-\oint_S dS_j\,\Pi^{aj}\,,
\label{15}
\end{equation}
where $\Pi^{aj}=-4ke\,\Sigma^{a0j}$. The expression above is the definition 
for the gravitational energy-momentum presented in ref. \cite{Maluf4}, 
obtained in the framework of the vacuum field equations in Hamiltonian form. 
It is invariant under coordinate transformations of the three-dimensional space
and under time reparametrizations. We note that (10) is a true energy-momentum 
conservation equation.

By substituting eq. (12) in the left hand side of eq. (11), and assuming that
the energy-momentum tensor for matter fields $T^{\lambda\mu}$ vanishes, which 
is the case for the Kerr space-time, we find

\begin{equation}
{{dP^a}\over {dt}}=
-\oint_S dS_j\,
\left[e\,e^a\,_\mu t^{j\mu} \right]\,.
\label{16}
\end{equation}
Considering now eq. (8), we rewrite the right hand side of the equation above
in the form

\begin{equation}
{{dP^a}\over {dt}}=
-4k\oint_S dS_j\,
\partial_\nu(e\Sigma^{a j\nu})\,.
\label{17}
\end{equation}
Now we restrict the Lorentz index $a$ to be $a=(i)$, where $i=1,2,3$, and 
write eq. (17) as

\begin{equation}
{{dP^{(i)}}\over {dt}}= -\oint_S dS_j\, \phi^{(i)j}
=-\oint_S dS_j \left[ e e^{(i)}\,_\mu t^{j\mu}\right] \,,
\label{18}
\end{equation}
where

\begin{equation}
\phi^{(i)j}=4k\partial_\nu(e\Sigma^{(i)j\nu}) \,.
\label{19}
\end{equation}
Equation (18) is precisely eq. (39) of Ref. \cite{Maluf1}.

The left hand side of eq. (18) represents the momentum of the field divided by
time, and therefore has the dimension of force. Since on the right hand side 
$dS_j$ is an element of area, we see that $-\phi^{(i)j}$ represents the  
pressure along the $(i)$ direction, over and element of area oriented along 
the $j$ direction. In cartesian coordinates the index $j=1,2,3$  represents the
directions $x,y,z$ respectively. 

In the next section we will consider the Kerr solution in terms of the
Boyer-Lindquist coordinates, which are spherical type coordinates.  In this 
case, we have $j=r,\theta,\phi$. In order to
obtain the radial pressure over the event horizon we need to consider only the
index $j=1$, which represents the radial direction.  Thus in spherical type 
coordinates we define  $-\phi^{(r)1}$ as

\begin{equation}
-\phi^{(r)1}=-(\sin\theta\cos\phi \,\phi^{(1)1}+
\sin\theta\sin\phi\,\phi^{(2)1} +\cos\theta\,\phi^{(3)1})\,,
\label{20}
\end{equation}
and from the expression above we define the radial pressure $p$ according to

\begin{equation}
p(r)=\int_0^{2\pi}d\phi \int_0^\pi d\theta \lbrack - 
\phi^{(r)1}  \rbrack\,.
\label{21}
\end{equation}
In view of eqs. (19) and (20), we see that $\phi^{(r)1}$ is a density.

In ref. \cite{Maluf1} we have written eq. (18) in vectorial form as

$$ {{d{\bf P}}\over {dt}}=\int _{\Delta S}\,d\theta d\phi \lbrack -\phi^{(r)1}
\rbrack \hat{\bf r}\,, $$
where $\Delta S$ is a spherical open surface of constant radius in the 
Schwarzschild
space-time with mass $m=GM/c^2$, according to eq. (44) of ref. \cite{Maluf1}.
Considering the surface $\Delta S$ to be $\Delta S=r^2 d\Omega$, where 
$d\Omega$ is a solid angle, we have shown that in the limit $r>>m$ the equation
above is simplified as 

$${d \over{ dt}}\biggl({{\bf P}\over M}\biggr)
=-{{GM}\over r^2} d\Omega\, \hat{\bf r}\,.$$

Finally, we note that all definitions presented in this section follow 
exclusively from the field equations (8).

\section{Radial pressure over the external event horizon of the Kerr
black hole}

In terms of the Boyer-Lindquist coordinates, the Kerr solution is given by

\begin{eqnarray}
ds^2&=&
-{{\psi^2}\over {\rho^2}}dt^2-{{2\chi\sin^2\theta}\over{\rho^2}}
\,d\phi\,dt
+{{\rho^2}\over {\Delta}}dr^2 \\ \nonumber
&{}&+\rho^2d\theta^2+ {{\Sigma^2\sin^2\theta}\over{\rho^2}}d\phi^2\,,
\label{22}
\end{eqnarray}
with the following definitions:

\begin{eqnarray}
\Delta&=& r^2+a^2-2mr\,,  \nonumber \\
\rho^2&=& r^2+a^2\cos^2\theta \,,  \nonumber \\
\Sigma^2&=&(r^2+a^2)^2-\Delta a^2\sin^2\theta\,,  \nonumber \\
\psi^2&=&\Delta - a^2 \sin^2\theta\,, \nonumber \\
\chi &=&2amr\,.
\label{23}
\end{eqnarray}

The tetrad field is obtained out of the metric tensor according to the 
relation $e^a\,_\mu e^b\,_\nu \eta_{ab}=g_{\mu\nu}$. The two properties that
may determine the tetrad field are the following. First, the tetrad field must
be adapted to a field of observers whose trajectories and velocities in 
space-time are given by $x^\mu(s)$ and $u^\mu(s)=dx^\mu/ds$, respectively,
where $s$ is the proper time. Then we identify $e_{(0)}\,^\mu =u^\mu$.
This equation fixes three conditions on the tetrad field: $e_{(0)}\,^i=u^i$.
And second, one may choose the unit vectors
$e_{(1)}\,^\mu$, $e_{(2)}\,^\mu$ and $e_{(3)}\,^\mu$ to be oriented 
asymptotically along
the unit cartesian vectors $\hat{\bf x}$, $\hat{\bf y}$, $\hat{\bf z}$.

Since the tetrad field is a kind of square root of the metric tensor, it may
not be defined in every region of the space-time. For instance, if we choose 
the observer to be static in space-time, then the tetrad field is
defined only in the region $r> r_+^*$, where 
$r_+^*=m+\sqrt{m^2-a^2\cos^2\theta}$ represents the external boundary
of the ergosphere of the Kerr space-time. This result is justified because 
inside the ergosphere it is not possible to  maintain any observer in 
static regime. Inside the ergosphere, all observers are necessarily dragged 
in circular motion
by the gravitational field.  The four-velocity of observers that circulate 
around the black hole inside the ergosphere, under the action of the 
gravitational field of the Kerr space-time, is given by 

\begin{equation}
u^\mu(t,r,\theta,\phi)
={{\rho \Sigma}\over{(\psi^2\Sigma^2+\chi^2\sin^2\theta)^{1/2}}}
(1,0,0,{\chi \over {\Sigma^2}})\,,
\label{24}
\end{equation}
where all functions are defined in eq. (23). It is possible to show that 
if we restrict the radial coordinate to $r=r_+$, the $\mu=3$ component of
eq. (24) becomes

$${\chi \over {\Sigma^2}}= {a\over {2mr_+}} =
{a \over {a^2+r_+^2}}= \Omega_H\,,$$
and $\Omega_H$ is the angular velocity of the external event
horizon of the Kerr space-time. The quantity

$$\omega(r)= -{{g_{03}}\over{g_{33}}}={{\chi}\over{ \Sigma^2}}$$
is the dragging velocity of inertial frames.

The tetrad field (i) that is adapted to observers whose four-velocities are 
given by eq. (24), i.e., for which $e_{(0)}\,^\mu =u^\mu$, 
and consequently is defined in the region $r\ge r_+$, 
(ii)  whose $e_{(i)}\,^\mu $ components in cartesian coordinates
are oriented along the unit vectors
$\hat{\bf x}$, $\hat{\bf y}$, $\hat{\bf z}$, and (iii) that is asympotically 
flat, is given by

\begin{equation}
e_{a\mu}=\pmatrix{-A&0&0&0\cr
B\sin\theta\sin\phi
&C\sin\theta\cos\phi& D\cos\theta\cos\phi&-E\sin\theta\sin\phi\cr
-B\sin\theta\cos\phi
&C\sin\theta\sin\phi& D\cos\theta\sin\phi& E\sin\theta\cos\phi\cr
0&C\cos\theta&-D\sin\theta&0}\,,
\label{25}
\end{equation}
where

\begin{eqnarray}
A&=& {{(g_{03}g_{03}-g_{00}g_{33})^{1/2}}\over{(g_{33})^{1/2}}}\,,
\nonumber \\
B&=&-{{ g_{03}}\over {(g_{33})^{1/2} \sin\theta}}\,, \nonumber \\
C&=&(g_{11})^{1/2}\,, \nonumber \\
D&=&(g_{22})^{1/2}\,, \nonumber \\
E&=& {{(g_{33})^{1/2}}\over {\sin\theta}}\,.
\label{26}
\end{eqnarray}
We emphasize that a relevant feature of the tetrad field above is that it is
defined from the spatial infinity up to the external event horizon of the Kerr
black hole. It is the unique configuration that satisfies the conditions stated
above, since six conditions are imposed on $e^a\,_\mu$. Therefore we may 
evaluate the gravitational pressure on the external event horizon.

Now we have all necessary quantities to evaluate $\phi^{(r)1}$ given by eq. 
(20). For this purpose we need eqs. (6), (25) and (26). The calculations are
rather lengthy, but otherwise straightforward. We find it relevant to show some
intermediate steps. First, we present the non-vanishing contributions to 
$\phi^{(r)1}(t,r,\theta\,\phi)$. They are given by

\begin{eqnarray}
\phi^{(r)1}&=&4k\cos\theta\,\partial_2(eC\cos\theta\Sigma^{112}-
eD\sin\theta\Sigma^{212}) \nonumber \\
&+& 4k\sin\theta\,\partial_2(eC\sin\theta\Sigma^{112}+
eD\cos\theta \Sigma^{212}) \nonumber \\
&+& 4k\,e\sin^2\theta(B\Sigma^{013}-E\Sigma^{313})\,,
\label{27}
\end{eqnarray}
recalling that $e$ is the determinant of $e^a\,_\mu$. In the expression of 
$\phi^{(r)1}$ we have neglected all terms that are linear in $\sin\phi$ and
$\cos\phi$, since by integration these terms will not contribute to $p$ given
by eq. (21). In order to evaluate the expression above we need the following
relations:

\begin{eqnarray}
\Sigma^{112}&=&{1\over 2} g^{11} g^{22}(-g^{00}T_{002}-g^{03}T_{302}
+g^{03}T_{023}+g^{33}T_{323})\,, \nonumber \\
\Sigma^{212}&=&{1\over 2} g^{11} g^{22}(g^{00}T_{001}+g^{03}T_{301}
-g^{03}T_{013}-g^{33}T_{313})\,, \nonumber \\
B\Sigma^{013}-E\Sigma^{313}&=&-{{g^{11}}\over{\sqrt{g_{33}}\sin\theta\,D}}
\biggl[ {1\over 4}g_{03}(-T_{301}+T_{013}+T_{103}) \nonumber \\
&{}&+{1\over 2}(g_{33}T_{001}-g^{22}\,D\,T_{212})\biggr]\,.
\label{28}
\end{eqnarray}
Finally, the torsion tensor components 
$T_{\lambda \mu\nu}=e^a\,_\lambda T_{a\mu\nu}$ are given by

\begin{eqnarray}
T_{001}&=&{1\over 2}\partial_1(A^2-B^2\sin^2\theta)\,, \nonumber \\
T_{301}&=&E\partial_1 B\sin^2\theta\,, \nonumber \\
T_{002}&=&{1\over 2}\partial_2(A^2-B^2\sin^2\theta)\,, \nonumber \\
T_{302}&=&E\partial_2 (B\sin^2\theta)\,, \nonumber \\
T_{103}&=&-BC\sin^2\theta\,, \nonumber \\
T_{212}&=&D\partial_1 D \cos^2\theta-DC\cos^2\theta-
D(\partial_2 C)\sin\theta \cos\theta\,, \nonumber \\
T_{013}&=&-B(\partial_1 E-C)\sin^2\theta\,, \nonumber \\
T_{313}&=& E(\partial_1 E -C)\sin\theta\,, \nonumber \\
T_{023}&=& -B(\partial_2 E)\sin^2\theta -
B(E-D)\sin\theta \cos\theta\,, \nonumber \\
T_{323}&=& E(\partial_2 E)\sin^2\theta + E(E-D)\sin\theta\cos\theta\,.
\label{29}
\end{eqnarray}

After a large number of calculations and simplifications we evaluate
expression (20), and arrive at

\begin{equation}
\phi^{(r)1}=2k(r_+-m){{\sin\theta} \over{\rho_+}}+
2k{{\sin\theta}\over{\rho_+^3}}\biggl[ma^2 \sin^2\theta+
2mr_+(r_+-m)\biggr]\,,
\label{30}
\end{equation}
where $\rho_+=\rho(r_+)$.

Let us analyse the case $a=J/m=0$ and obtain the radial pressure over the
event horizon of the  Schwarzschild black hole. By making $a=0$ we obtain
$\phi^{(r)1}=2k\sin\theta$, when $r_+=2m$. Applying this expression in
eq. (21) yields $p=-c^3/2G$. The latter is the gravitational pressure
over the event horizon of the Schwarzschild black hole. The negative sign 
means that the pressure is exerted towards the centre of the black hole.
Recall that $\phi^{(r)1}$ is a density. Therefore it incorporates a quantity
of the type $r^2\sin\theta$. 

In the limit $a\rightarrow 0$, eq. (27) reduces to 

$$\phi^{(r)1}=2k\,\sin\theta\biggl[ {{2m}\over r}(1-e^\lambda)-
e^\lambda(1-e^\lambda)^2\biggr]\,,$$
where $e^\lambda=\sqrt{-g_{00}}$.  This expression was obtained in ref. 
\cite{Maluf1}, in the context of the Schwarzschild black hole. We note, 
however, that in eqs. (43) and (44) for $\phi^{(r)1}$ in ref. \cite{Maluf1},
the quantity $e^\lambda$, that multiplies the term $(1-e^\lambda)^2$ on the 
right hand side of the expression above, is missing.

The radial pressure over the external event horizon of the Kerr black hole
is obtained by integration of the angular variables in eq. (30), according 
to eq. (21), and making $r=r_+$. The integration is not difficult, and
eventually we obtain

\begin{equation}
p={1\over 4}\bigg[-{{4m}\over{\sqrt{2mr_+}}}+{{(2m-r_+)}\over a}
\ln \biggl( {{\sqrt{2mr_+}+a}\over{\sqrt{2mr_+}-a}}\biggr)\biggr]\,.
\label{31}
\end{equation}
It is not difficult to see that the value of the pressure is negative. This 
is the expression that we will use in the next section in the analysis of the 
thermodynamics of the Kerr black hole.

\section{The thermodynamic relation for $TdS$}

In this section, we will consider a Kerr black hole of mass $m$ and angular
momentum $J$ in stationary state. Then by means of the Penrose process we 
consider that $m$ and $J$ vary according to $m \rightarrow m + dm$ and 
$J \rightarrow J + dJ$. The standard thermodynamic relation 
for black holes is given by 
eq. (1). We will obtain here the thermodynamic relation $TdS$ entirely within
the framework of the TEGR, with no identification between $TdS$ and the 
variation $dA$ of the area of the event horizon.
Since the first law of thermodynamics is given by
$TdS= dE + pdV$, we need to calculate separately the variations $dE$ and $pdV$
in the Penrose process.  Let us begin with $dE$.

In ref. \cite{Maluf4} we have calculated the energy contained within the 
external event horizon of the Kerr black hole, making use of the tetrad field
given by eqs. (25) and (26). The latter is precisely equal to the set of tetrad
fields given by expression (4.9)
of ref. \cite{Maluf4}. The resulting value for the gravitational energy 
contained within a surface of constant radius $r=r_+$ is given by eq. (5.4) of
the latter reference. It reads
\begin{equation}
E=m\biggl[ {\sqrt{2p}\over 4}+{{6p-\lambda^2}\over {4\lambda}}
\ln \biggl({{\sqrt{2p}+\lambda}\over p}\biggl) \biggr]\,,
\label{32}
\end{equation}
where

$$p=1+\sqrt{1-\lambda^2}\,, \ \ \ \ \ a=\lambda m\,, \ \ \ \ \ 
0\le \lambda \le 1\,.$$
The parameter $p$ used above (and in ref. \cite{Maluf4}) is not to be confused 
with radial pressure $p$. 
The remarkable feature of the expression above is that, for any value of the
parameter $\lambda$, it is strikingly close to $2m_{irr}$. We recall that the
energy contained within the event horizon of the Schwarzschild black hole of
mass $m$ is given by $2m$. For the Kerr black hole one expects that the
energy contained within the external event horizon is given by $2m_{irr}$, 
which is the amount of gravitational energy that cannot be extracted from the
black hole in the Penrose process.
An analysis of the various gravitational energy expressions for the 
Schwarzschild and Kerr black holes has been made in ref. \cite{Bergqvist}. It 
was concluded that none of the known expressions yield the value $2m_{irr}$
for the Kerr black hole, but all of them yield $2m$ for the Schwarzschild black
hole. The almost coincidence between our expression and
$2m_{irr}$, for any value of $\lambda$, is displayed by Fig. 1 of
\cite{Maluf4}. Rewriting $E$ given by eq. (32) in terms of $(m,a)$ only we have

\begin{equation}
E={m\over 4}\biggl[ \sqrt{2r_+ \over m}   +{{(6m r_+ -a^2)}\over {ma}}
\ln\biggl( {{\sqrt{2mr_+} +a} \over r_+} \biggr) \biggr]\,,
\label{33}
\end{equation}

We will obtain $dE$ and write it in terms of the differentials $dm$ and
$dJ$, which are the differentials that arise in eq. (1). Thus we will 
calculate 

$$dE= {{\partial E}\over {\partial m}}dm + 
{{\partial E}\over {\partial J}}dJ\,.$$
Recall that $r_+=m+\sqrt{m^2-J^2/m^2}$. The calculation is rather lengthy. 
We obtain

\begin{eqnarray}
dE&=&{1\over 4} \sqrt{{2r_+}\over m}dm 
+{1\over 8}\biggl({{6r_+}\over a}-{a\over m}\biggr)
\ln \biggl( {{\sqrt{2mr_+} +a} \over{ \sqrt{2mr_+ -a}}}\biggr)dm \nonumber \\
&+&{m\over 8} \biggl( {{12r_+}\over{a\sqrt{m^2-a^2}}}+{{2a}\over m^2} \biggr)
\ln \biggl( {{\sqrt{2mr_+} +a} \over{ \sqrt{2mr_+ -a}}}\biggr)dm \nonumber \\
&+&{{a^2-3mr_+}\over \sqrt{2mr_+(m^2-a^2)}}dm\nonumber \\
&-&{m\over 8}\biggl({{6r_+}\over {a^2\sqrt{m^2-a^2}}}+{1\over m^2}\biggr)
\ln \biggl( {{\sqrt{2mr_+} +a} \over{ \sqrt{2mr_+ -a}}}\biggr) dJ\nonumber \\
&+&\biggl({{3mr_+}\over {2a}}-{a\over 2}  \biggr)
{1\over \sqrt{2mr_+(m^2-a^2)}}dJ\,.
\label{34}
\end{eqnarray}

Next we will calculate $pdV$. Since $\phi^{(r)1}$ is a density, the 
differential $pdV$ is evaluated as

\begin{equation}
pdV=\biggl[ \int_S(-\phi^{(r)1})\,d\theta d\phi \biggr]dr_+
=p\,dr_+\,,
\label{35}
\end{equation}
where $S$ is the surface of constant radius $r=r_+$, and $p$ is given by
eq. (31). The differential $dr_+$ is given by

\begin{equation}
dr_+={{(r_+ + a^2/m)}\over \sqrt{m^2-a^2}}dm-
{{a/m}\over\sqrt{m^2-a^2}}dJ\,.
\label{36}
\end{equation}
Since we are assuming that $dr_+$, $dm$ and $dJ$ are infinitesimals, the 
present analysis is not valid when the squere root $\sqrt{m^2-a^2}$ approaches
zero, i.e.,  when $a$ is very close to $m$. After a number of calculations and
simplifications we obtain

\begin{eqnarray}
pdV&=&-{{(mr_+ +a^2)}\over \sqrt{2mr_+(m^2-a^2)}} dm \nonumber \\
&+&{1\over 4}\biggl( {{2a}\over\sqrt{m^2-a^2}} -{a\over m}\biggr)
\ln \biggl({{\sqrt{2mr_+}+a}\over{\sqrt{2mr_+}-a}}\biggr) dm  \nonumber \\
&+&{ a \over {\sqrt{2mr_+(m^2-a^2)} }}dJ \nonumber \\
&-& {1\over 4}\, {{\lbrack 2-(r_+/m)\rbrack}\over \sqrt{m^2-a^2}}
\ln\biggl({{\sqrt{2mr_+}+a}\over{\sqrt{2mr_+}-a}}\biggr)dJ\,.
\label{37}
\end{eqnarray}

We are now in a position to write the first law of thermodynamics 
$TdS=dE + pdV$ out of eqs. (34) and (37). We will combine the latter equations
and write $TdS$ in the form

\begin{equation}
TdS=f_1(m,a)dm +f_2(m,a) dJ\,.
\label{38}
\end{equation}
The functions $f_1$ and $f_2$ are obtained by means of straightforward algebra.
They read

\begin{eqnarray}
f_1(m,a)&=&{r_+ \over {{2\sqrt{m^2-a^2}}}} 
\biggl[ \biggl( {{r_+ -9m \over\sqrt{2mr_+}}}\biggr) \nonumber \\
&+& \biggl( {{r_+^2 + mr_+ + 16m^2}\over {4ma}}\biggr)
\ln\biggl({{\sqrt{2mr_+}+a}\over{\sqrt{2mr_+}-a}}\biggr)
\biggr] \,, \\
f_2(m,a)&=&{{r_+}\over {2a}}\biggl[ \biggl({3\over {\sqrt{2mr_+}}}
\biggr) \nonumber \\
&+&\biggl( {{3r_+^2-11mr_++4m^2}\over{4ma\sqrt{m^2-a^2}}}
\biggr)\ln\biggl({{\sqrt{2mr_+}+a}\over{\sqrt{2mr_+}-a}}\biggr)\biggr]\,.
\label{39,40}
\end{eqnarray}

Expressions (38), (39) and (40) should be compared with the standard expression
for $TdS$ given by eq. (1), which may be written as

\begin{equation}
TdS=f_3(m,a)dm + f_4(m,a)dJ\,,
\label{41}
\end{equation}
where

\begin{eqnarray}
f_3(m,a)&=& 1 \,, \\
f_4(m,a)&=& -\Omega_H = -{a\over {2mr_+}}\,.
\label{42,43}
\end{eqnarray}

In order to compare eqs. (38-40) and (41-43), we will plot in the same graph 
the functions $f_1$ and $f_3$, and $f_2$ and $f_4$. For this purpose, we 
choose three arbitrary values of the parameter $\lambda$, where $a=\lambda m$.
We have chosen $\lambda =0.2$, $\lambda=0.5$ and $\lambda =0.9$.  As we 
explained earlier, the values of $\lambda$ close to $1$ are not allowed in
the analysis, since in this case the differential $dr_+$ is no longer an
infinitesimal. All curves in all graphs are functions of the mass parameter 
$m$.

In all figures, the 
dotted lines represent $f_3$ and $f_4$, namely, they represent the standard 
thermodynamic relation given by eq. (1). The regular (continuous) 
lines represent our result
given by eqs. (38-40). The main results are that (i) the qualitative behaviour 
of $f_1$ and $f_2$ is the same in all figures,  and (ii) $f_1$ and 
$f_2$ are qualitatively similar to $f_3$ and $f_4$, respectively, in all 
figures. At first sight it is
surprising that $f_1$ yields a horizontal line, and that $f_2$ looks like 
$f_4$. We will explain this qualitative behaviour in the next section. 
Our first conclusion
is that the thermodynamic relation given by eq. (38), obtained entirely out of
the definitions in the TEGR, is realistic and comparable to eq. (1).
\bigskip

\begin{figure}[htb]
\begin{minipage}[t]{0.45\linewidth}
\includegraphics[width=\linewidth]{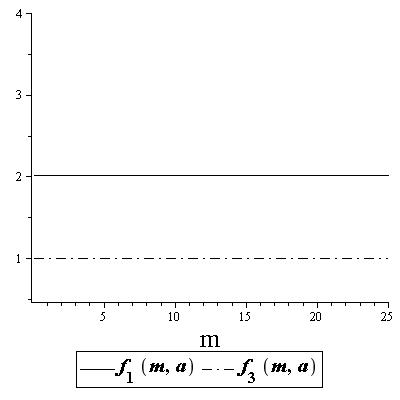}
\end{minipage} \hfill
\begin{minipage}[t]{0.45\linewidth}
\includegraphics[width=\linewidth]{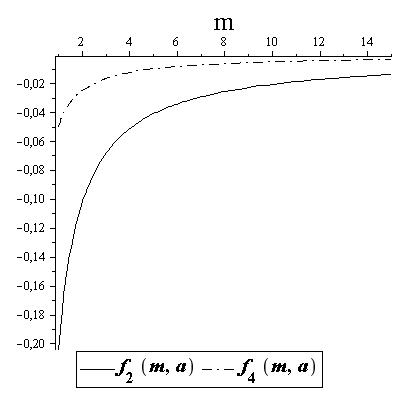}
\end{minipage}
\caption{$\lambda=0.2$}
\end{figure}
\bigskip
\begin{figure}[htb]
\begin{minipage}[t]{0.45\linewidth}
\includegraphics[width=\linewidth]{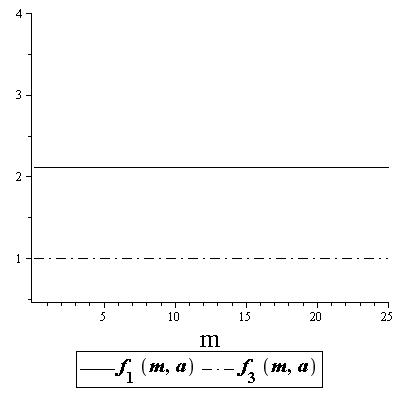}
\end{minipage} \hfill
\begin{minipage}[t]{0.45\linewidth}
\includegraphics[width=\linewidth]{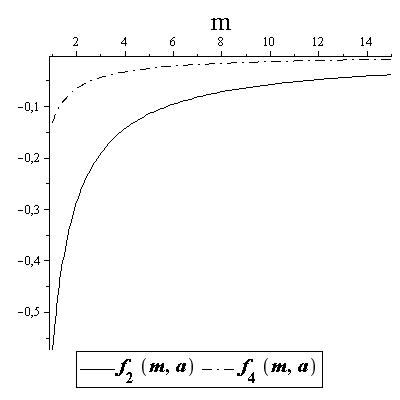}
\end{minipage}
\caption{$\lambda=0.5$}
\end{figure}

\bigskip
\begin{figure}[htb]
\begin{minipage}[t]{0.45\linewidth}
\includegraphics[width=\linewidth]{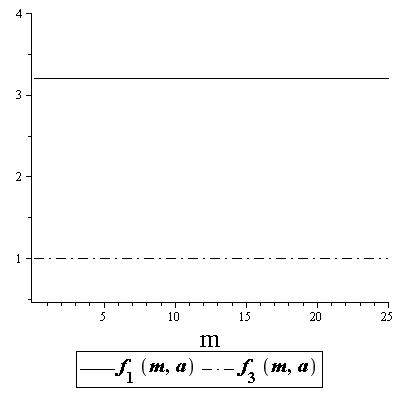}
\end{minipage} \hfill
\begin{minipage}[t]{0.45\linewidth}
\includegraphics[width=\linewidth]{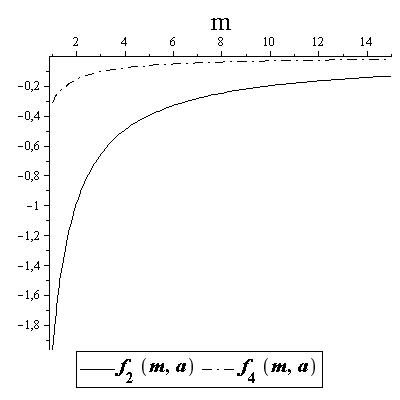}
\end{minipage}
\caption{$\lambda=0.9$}
\end{figure}

In the standard treatment of the thermodynamics of black holes one, does not
make use of the concept of pressure. We find it appropriate to analyse the
thermodynamic relation for $TdS$, when the latter is due the variation of 
the energy only. Thus let us analyse

\begin{equation}
TdS=dE \equiv f_{0m}(m,a) dm + f_{0J}(m,a) dJ\,.
\label{44}
\end{equation}
In view of eq. (34) we find

\begin{eqnarray}
f_{0m}(m,a)&=&
-\frac{r_+}{2\sqrt{m^2-a^2}}\biggl[ \left(\frac{r_++3m}{\sqrt{2mr_+}}\right)
\nonumber \\
&+&\left(\frac{r_+^2-9mr_+-4m^2}{4ma}\right)
\ln{\left(\frac{\sqrt{2mr_+}+a}{\sqrt{2mr_+}-a}\right)}\biggr]\,,
\label{45}
\end{eqnarray}
and

\begin{equation}
f_{0J}(m,a)=\frac{r_+(r_+
+m)}{2a\sqrt{m^2-a^2}}\left[\left(\frac{1}{\sqrt{2mr_+}}\right)
+\left(\frac{r_+ -4m}{4ma}\right)
\ln{\left(\frac{\sqrt{2mr_+}+a}{\sqrt{2mr_+}-a}\right)}\right]\,.
\label{46}
\end{equation}

\begin{figure}[htb]
\begin{minipage}[t]{0.45\linewidth}
\includegraphics[width=\linewidth]{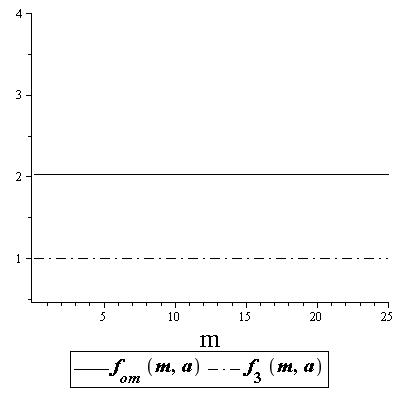}
\end{minipage} \hfill
\begin{minipage}[t]{0.45\linewidth}
\includegraphics[width=\linewidth]{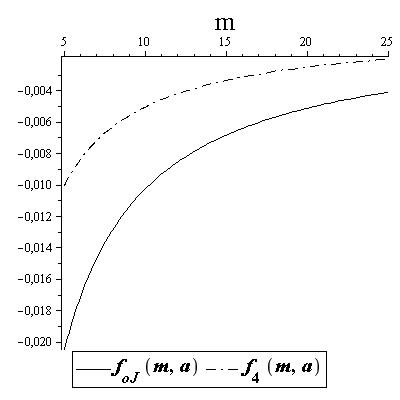}
\end{minipage}
\caption{$\lambda=0.2$}
\end{figure}

\begin{figure}[htb]
\begin{minipage}[t]{0.45\linewidth}
\includegraphics[width=\linewidth]{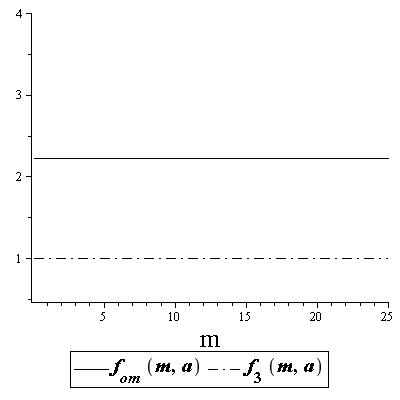}
\end{minipage} \hfill
\begin{minipage}[t]{0.45\linewidth}
\includegraphics[width=\linewidth]{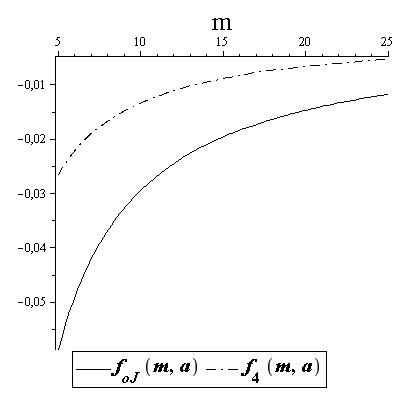}
\end{minipage}
\caption{$\lambda=0.5$}
\end{figure}
\bigskip

\begin{figure}[htb]
\begin{minipage}[t]{0.45\linewidth}
\includegraphics[width=\linewidth]{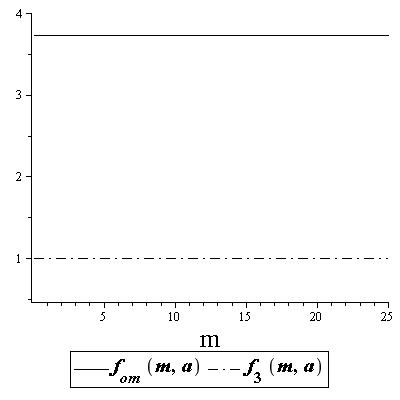}
\end{minipage} \hfill
\begin{minipage}[t]{0.45\linewidth}
\includegraphics[width=\linewidth]{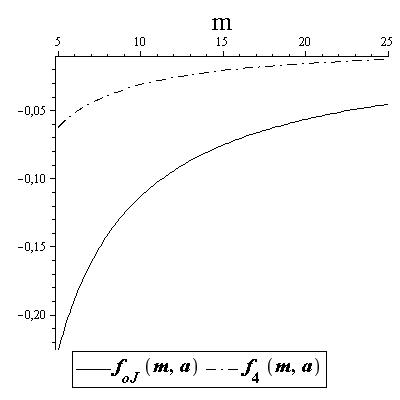}
\end{minipage}
\caption{$\lambda=0.9$}
\end{figure}

We need to compare $f_{0m}$ with $f_3$, and $f_{0J}$ with $f_4$. Again
we chose the values $0.2\;$, $0.5\;$ and $0.9$ for $\lambda=a/m$, and
therefore all curves will be functions of the parameter $m$. Once again
we observe that $f_{0m}$ is qualitatively similar to $f_3$, and that
$f_{0J}$ is qualitatively similar to $f_4$.

An immediate conclusion about these figures is the following. By comparing the
horizontal lines in Fig. 3 (with pressure, $\lambda =0.9$) 
and Fig. 6 (with no pressure, $\lambda=0.9$) we 
observe that the numerical, constant value of the straight line is lower in
Fig. 3 compared to Fig. 6. Admitting that the term $f_1 dm$ overweights 
the effect of the term $f_2 dJ$ in the Penrose process 
(since the particle may have vanishing angular momentum), then we see that
the effect of the presence of the gravitational
pressure is to increase the efficiency of
the Penrose process. The reason is that the smaller the
value of $dE$, more energy can be extracted from the black hole. However, by
inspecting figures 1, 2 and 3 
we see that the efficiency of the Penrose process is 
lower in the thermodynamic process dictated by relation (38), compared with 
the standard relation (1) or (41), since after the variations 
$m  \rightarrow m + dm$ and $J  \rightarrow J + dJ$, the energy variation 
$dE$ is higher in the context of the present analysis, namely, in the context
of eq. (38). The higher is $dE$, less efficient is the Penrose process.

\section{ The dependence of the figures on the parameter $m$}

In this section, we will explain the reason for the similarity between 
$f_1$, $f_3$ and $f_{0m}$, and between $f_2$, $f_4$ and $f_{0J}$. The idea is
very simple. We just substitute $\lambda m$ for $a$ in all expressions, and
factorize the mass parameter $m$. After this procedure we see that $f_1$ and
$f_{0m}$ depend only on $\lambda$, i.e., they are independent of $m$, and 
that $f_2$, $f_{0J}$ and $f_4$ all depend on $1/m$ times a function that 
depends only on $\lambda$. We find

\begin{eqnarray}
f_1(m,\lambda)&=&\biggl( {{1+\sqrt{1-\lambda^2}}\over {2\sqrt{1-\lambda^2}}}
\biggr) \biggl\{ {{\sqrt{1-\lambda^2}-8}\over \sqrt{2+2\sqrt{1-\lambda^2}}}
\nonumber \\
&+& \biggl( {{19+3\sqrt{1-\lambda^2}-\lambda^2}\over {4\lambda}} \biggr)
\ln \biggl[ 
{{ \sqrt{2+2\sqrt{1-\lambda^2}} +\lambda} \over 
 { \sqrt{2+2\sqrt{1-\lambda^2}} -\lambda}} 
\biggr] \biggr\} \,,
\label{47}
\end{eqnarray}

\begin{eqnarray}
f_2(m,\lambda)&=&
{1\over m}\biggl( {{1+\sqrt{1-\lambda^2}}\over {2\lambda}}
\biggr) \biggl\{ {{3}\over \sqrt{2+2\sqrt{1-\lambda^2}}}
\nonumber \\
&-& \biggl( {{ 1+5\sqrt{1-\lambda^2} + 
3\lambda^2}\over { 4\lambda\sqrt{1-\lambda^2}}} \biggr)
\ln \biggl[ 
{{ \sqrt{2+2\sqrt{1-\lambda^2}} +\lambda} \over 
 { \sqrt{2+2\sqrt{1-\lambda^2}} -\lambda}} 
\biggr] \biggr\} \,,
\label{48}
\end{eqnarray}

\begin{eqnarray}
f_{0m}(m,\lambda)&=&
-\biggl( {{1+\sqrt{1-\lambda^2}}\over {2\sqrt{1-\lambda^2}}}
\biggr) \biggl\{ {{\sqrt{1-\lambda^2}+4}\over \sqrt{2+2\sqrt{1-\lambda^2}}}
\nonumber \\
&-& \biggl( {{11+7\sqrt{1-\lambda^2}+\lambda^2}\over {4\lambda}} \biggr)
\ln \biggl[ 
{{ \sqrt{2+2\sqrt{1-\lambda^2}} +\lambda} \over 
 { \sqrt{2+2\sqrt{1-\lambda^2}} -\lambda}} 
\biggr] \biggr\} \,,
\label{49}
\end{eqnarray}

\begin{eqnarray}
f_{0J}(m,\lambda)&=&
{1\over m}\biggl( {{(1+\sqrt{1-\lambda^2})(2+\sqrt{1-\lambda^2})}\over 
{2\lambda \sqrt{1-\lambda^2}}}
\biggr) \biggl\{ {{ 1}\over \sqrt{2+2\sqrt{1-\lambda^2}}}
\nonumber \\
&+& \biggl( {{ \sqrt{1-\lambda^2} - 
3 }\over { 4\lambda }} \biggr)
\ln \biggl[ 
{{ \sqrt{2+2\sqrt{1-\lambda^2}} +\lambda} \over 
 { \sqrt{2+2\sqrt{1-\lambda^2}} -\lambda}} 
\biggr] \biggr\} \,,
\label{50}
\end{eqnarray}

\begin{equation}
f_4(m,\lambda)=-{\lambda \over {2m(1+\sqrt{1-\lambda^2})}}\,.
\label{51}
\end{equation}
Thus we see that $f_1$ and $f_{0m}$ are independent of $m$, and 
$f_2$, $f_{0J}$ and $f_4$ depend on $1/m$. These features explain the shape
of the curves in the graphs.

\section{Conclusions and summary of the results}

In this article, we have investigated the thermodynamics of the Kerr black hole
in the context of the teleparallel equivalent of general relativity. We have 
considered: {\bf (i)} a stationary black hole with mass and angular momentum 
per unit mass $(m,a=J/m)$; {\bf (ii)} the Penrose process by means of which 
$m$ and $J$ vary according to $m\rightarrow m+dm$, $J\rightarrow J +dJ$;
{\bf (iii)} the expressions for gravitational energy and pressure, $E$ and
$p$ respectively, that arise in teleparallel equivalent of general relativity.
In particular, we have considered the expression for $E$ given by eq. (33). 
The numerical values of the latter, for specific values of $m$ and $a$, are
strikingly close to $2m_{irr}$, as shown in ref. \cite{Maluf4}. With all
these quantities we evaluated $TdS = dE+pdV$.

There are two main results in the article. First, we have evaluated the 
gravitational pressure over the external event horizon of the black hole. The
radial pressure, with no dependence on the angular variables, is given by eqs.
(21) and (31). The second result is the thermodynamic relation (38), with
the definitions (39) and (40). This relation is obtained without any 
consideration of the area of the external event horizon of the black hole. It
follows entirely from the definitions that arise in the TEGR. Although 
relations (38-40) look like quite different from the standard thermodynamic
relation (41-43), we see from figures 1-3 that both thermodynamic relations
are qualitatively similar. Note, however, that in the present analysis the 
variations of the gravitational energy and pressure that occur in view of the 
Penrose process are necessarily subject to the conservation equation (11).
Admitting that the higher is $TdS$, the lower is
the efficiency of the Penrose process, then it seems that in the context of
the TEGR the Penrose process is less efficient than in the standard process.
Finally, by analysing Figs. 3-6 we see that the presence of the gravitational
pressure in the thermodynamic relation for $TdS$ does not imply a significant
change. We believe that both the concept of gravitational pressure
and our thermodynamic relation $TdS=dE + pdV$ given by eq. (38) are physically
consistent and realistic results.

\end{document}